\documentclass{article}
\usepackage{slashed,epsfig,amsmath,amssymb,enumitem}
\usepackage[papersize={8.5in,11in}]{geometry}
\usepackage{graphicx}
\begin{document}
\title{Modified Gravity Black Holes and their Observable Shadows}
\author{J. W. Moffat\\~\\
Perimeter Institute for Theoretical Physics, Waterloo, Ontario N2L 2Y5, Canada\\
and\\
Department of Physics and Astronomy, University of Waterloo, Waterloo,\\
Ontario N2L 3G1, Canada}
\maketitle




\begin{abstract}
The shadows cast by non-rotating and rotating modified gravity (MOG) black holes are determined by the two parameters mass $M$ and angular momentum $J=Ma$. The sizes of the shadows cast by the spherically symmetric static Schwarzschild-MOG and Kerr-MOG rotating black holes increase significantly as the free parameter $\alpha$ is increased from zero. The Event Horizon Telescope (EHT) shadow image measurements can determine whether Einstein's general relativity is correct or whether it should be modified in the presence of strong gravitational fields.
\end{abstract}

\maketitle

\section{Introduction and field equations}

The static spherically symmetric vacuum solution describing the final stage of the collapse of a body in terms of an enhanced gravitational constant $G$ and a gravitational repulsive force with a charge $Q=\sqrt{\alpha G_N}M$, has been derived in a modified gravitational theory (MOG)~\cite{Moffat,Moffat2}, where $\alpha$ is a parameter defined by $G=G_N(1+\alpha)$, and where $G_N$ is Newton's constant and $M$ is the total mass of the black hole. In the following, the optical shadows (silhouettes) cast by rotating and non-rotating black holes are determined.

The modified gravitational field equations are given by
\begin{equation}
\label{MOGgraveqs}
G_{\mu\nu}=-8\pi G T_{\phi\mu\nu}.
\end{equation}
The canonical energy-momentum tensor of matter $T_{M\mu\nu}$ in the gravitational field equations has been set equal to zero, and $G_{\mu\nu}=R_{\mu\nu}-(1/2)g_{\mu\nu}R$ is the Einstein tensor constructed from the Riemann tensor and its contractions. Moreover,
\begin{equation}
\label{Tphi}
T_{\phi\mu\nu}=-\frac{1}{4\pi}({B_\mu}^\sigma B_{\nu\sigma}-\frac{1}{4}g_{\mu\nu}B^{\sigma\beta}B_{\sigma\beta}),
\end{equation}
where $B_{\mu\nu}=\partial_\mu\phi_\nu-\partial_\nu\phi_\mu$ and where $\phi_\mu$ is the vector field with the source charge, $Q=\sqrt{\alpha G_N}M$. We also need the vacuum field equations:
\begin{equation}
\label{Bequation}
\nabla_\nu B^{\mu\nu}=\frac{1}{\sqrt{-g}}\partial_\nu(\sqrt{-g}B^{\mu\nu})=0,
\end{equation}
and
\begin{equation}
\label{Bcurleq}
\nabla_\sigma B_{\mu\nu}+\nabla_\mu B_{\nu\sigma}+\nabla_\nu B_{\sigma\mu}=0,
\end{equation}
where $\nabla_\nu$ is the covariant derivative with respect to the metric tensor $g_{\mu\nu}$.

\section{Schwarzschild-MOG and Kerr-MOG black holes}

The static spherically symmetric Schwarzschild-MOG metric obtained from the field equations (\ref{MOGgraveqs}) is given by
\begin{equation}
ds^2=\biggl(1-\frac{2G_N(1+\alpha)M}{r}+\frac{G_N^2\alpha(1+\alpha) M^2}{r^2}\biggr)dt^2-\biggl(1-\frac{2G_N(1+\alpha)M}{r}+\frac{G_N^2\alpha(1+\alpha)M^2}{r^2}\biggr)^{-1}dr^2-r^2d\Omega^2,
\end{equation}
where the numerator of the third term in parenthesis is $GQ^2=G_N^2\alpha(1+\alpha)M^2$~\cite{Moffat2} and $d\Omega^2=d\theta^2+\sin^2\theta d\phi^2$.

The Kerr-MOG metric inferred from our gravitational field equations has the form in Boyer-Lindquist coordinates $r,\theta,\phi$:
\begin{equation}
ds^2=\biggl(1-\frac{r_sr-r_Q^2}{\rho^2}\biggr)dt^2-\biggl[r^2+a^2+a^2\sin^2\theta\biggl(\frac{r_gr-r_Q^2}{\rho^2}\biggr)\biggr]\sin^2\theta d\phi^2
$$ $$
+2\sin^2\theta\biggl(\frac{r_gr-r_Q^2}{\rho^2}\biggr)dtd\phi-\frac{\rho^2}{\Delta}dr^2-\rho^2d\theta^2,
\end{equation}
where
\begin{equation}
\label{nonsingrho}
\rho^2=r^2+a^2\cos^2\theta.
\end{equation}
Moreover, $r_s=2G_N(1+\alpha)M$, $a=J/M$, $r_Q^2=G_N^2\alpha(1+\alpha)M^2$, where $J$ is the spin angular momentum and
\begin{equation}
\Delta=r^2-r_gr+a^2+r_Q^2.
\end{equation}
In the case of the Kerr-MOG metric the Kretschmann scalar invariant $R^{\mu\nu\sigma\beta}R_{\mu\nu\sigma\beta}$ is singular for $\rho^2=r^2+a^2\cos^2\theta=0$ and as in the case of the Kerr-Newman~\cite{Newman} black hole the singularity takes the form of a ring for $r=0,\theta=\pi/2$~\cite{Carter}.

Horizons are determined by the roots of $\Delta=0$:
\begin{equation}
r_\pm=G_N(1+\alpha)M\biggl[1\pm\sqrt{1-\frac{a^2}{G_N^2(1+\alpha)^2M^2}-\frac{\alpha}{1+\alpha}}\biggr].
\end{equation}
An ergosphere horizon is determined by $g_{00}=0$:
\begin{equation}
r_E=G_N(1+\alpha)M\biggl[1+\sqrt{1-\frac{a^2\cos^2\theta}{G_N^2(1+\alpha)^2M^2}-\frac{\alpha}{1+\alpha}}\biggr].
\end{equation}

\section{Shadows (silhouettes) of black holes}

We shall take it as given that our Schwarzschild-MOG and Kerr-MOG black holes are characterized by only the two parameters mass $M$ and angular momentum $J$. They are stationary and asymptotically flat solutions and they satisfy the ``no-hair'' theorem. An interesting consequence of these properties of the solution is that the shadow outline created by the black hole is determined by $M$ and $a=J/M$ and the relative position of an asymptotic observer. In the near future, it is expected that observations by the Event Horizon Telescope (EHT) can observe characteristic features of a black hole by the shapes of the shadows cast by the black hole~\cite{Bardeen,Luminet,deVries,deVries2}.

The apparent shape of a black hole and its outline is determined by geometrical optics. A electromagnetic wave propagates approximately on a congruence of light rays perpendicular to the wave fronts, bent by the curved spacetime geometry. For a black hole with a horizon $r_+$ the geometric approximation is valid only if the wavelength $\lambda$ is small compared to the typical radius of the spacetime curvature measured in a local patch. In Fig.1, we depict the curved light rays reaching an observer at infinity as if they are emitted from different directions.

The black hole casts a shadow in front of an illuminated background in the asymptotically flat region and the shadow is determined by a set of closed photon orbits. A photon moving on a closed orbit with radius $r$ in our Kerr-MOG spacetime with nonzero $a$ and $M$ has the apparent position in the (x,y) reference frame of a distant observer located in the angle of latitude $\theta$ (see Fig.1). The $x$ and $y$ coordinates are given by~\cite{deVries,deVries2}:
\begin{equation}
\label{xy}
x=\frac{r\Delta+r\alpha G_N^2(1+\alpha)M^2-G_N(1+\alpha)M(r^2-a^2)}{a[r-G_N(1+\alpha)M]\sin\theta},
$$ $$
y=\biggl\{\frac{4r^2\Delta}{[r-G_N(1+\alpha)M]^2}-(x+a\sin\theta)^2\biggr\}^{1/2},
\end{equation}
where
\begin{equation}
\Delta=r^2-2G_N(1+\alpha)Mr+a^2+\alpha G_N^2(1+\alpha)M^2.
\end{equation}

The variable $x$ in (\ref{xy}) is a monotonic function of $r$ for $r > G_N(1+\alpha)M$. By inversion we obtain the formula:
\begin{equation}
r=G_N(1+\alpha)M+\sqrt[3]{v+\sqrt{v^2+w^3}}+\sqrt[3]{v-\sqrt{v^2+w^3}},
\end{equation}
where
\begin{equation}
v=G_N(1+\alpha)M[G_N^2M^2(1+\alpha)-a^2],
\end{equation}
and
\begin{equation}
w=\frac{1}{3}[a^2-G_N^2M^2(3+\alpha)(1+\alpha)-ax\sin\theta].
\end{equation}\\

\
\begin{figure}
\centering \includegraphics[scale=0.4]{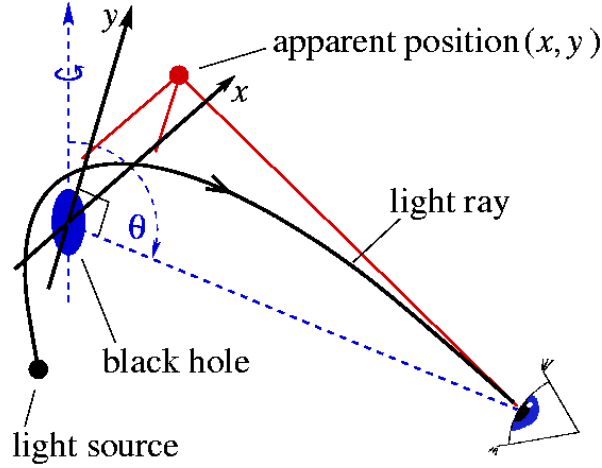}\\
\caption{\label{fig:shadow}The apparent position of a light ray with respect to the observer's projection plane in the x,y coordinates containing the center of the spacetime: x denotes the apparent distance from the rotation axis, and y the projection of the rotation axis itself (dashed line). The angle $\theta$ denotes the angle of latitude, reaching from the north pole at $\theta = 0$ to the south pole at $\theta = \pi$ (image by de Vries).}
\end{figure}

For the non-rotating case $a=0$ we have
\begin{equation}
\Delta=r^2-2G_N(1+\alpha)Mr+\alpha G_N^2(1+\alpha)M^2,
\end{equation}
and
\begin{equation}
x^2+y^2=\frac{r^4}{\Delta}.
\end{equation}
The size of the photosphere is determined by~\cite{Moffat2}:
\begin{equation}
\label{requation}
r_\gamma=r=\frac{3}{2}G_N(1+\alpha)M\biggl(1+\sqrt{1-\frac{8\alpha }{9(1+\alpha)}}\biggr).
\end{equation}

The shadow radius is given by
\begin{equation}
r_{\rm shad}\equiv\sqrt{x^2+y^2}=\frac{r^2}{\Delta^{1/2}},
\end{equation}
where the closed photon orbit radius $r$ is given by (\ref{requation}). We obtain for the shadow radius:
\begin{equation}
\label{shadowradius}
r_{\rm shad}=\frac{\left[3(1+\alpha)\pm\sqrt{(9+\alpha)(1+\alpha)}\right]^2}{\biggl\{4\left[(1+\alpha)\pm\sqrt{(9+\alpha)(1+\alpha)}\right]^2-16(1+\alpha)\biggr\}^{1/2}}G_NM.
\end{equation}

The shadow radius can be approximated by the linear expression: $r_{\rm shad}\sim (2.59+2\alpha)r_s$. We see that as $\alpha$ increases from the Schwarzschild black hole shadow radius with $\alpha=0$, the size of the MOG black hole shadow increases. The galaxy rotation curves and the dynamics of galaxy clusters were fitted with the value $\alpha=8.98\pm 0.34$~\cite{MoffatRahvar,MoffatRahvar2}. For $\alpha=9$ the shadow radius obtained from (\ref{shadowradius}) is $r_{\rm shad}=22.68r_s$. The effect of the $M^2$ contribution in the Kerr-MOG shadow is to decrease the distortion of the circular shadow for $a\neq 0$ in the Kerr black hole shadow. The significant increase in the shadow radius for the Schwarzschild-MOG and Kerr-MOG black holes as $\alpha$ becomes large compared to the Schwarzschild and Kerr black hole values, should be measurable when the shadow image data obtained by the EHT observations become available. We note that the value $\alpha=8.89$ used to fit the galaxy rotation curve and cluster data may not be applicable to the strong gravitational fields associated with the black holes in Sagittarius $A^*$ and M87.

\section{Conclusions}

We have investigated the black holes predicted by modified gravity theory for both rotating and non-rotating black holes.
The black hole shadows (silhouettes) against a bright background around super-massive black hole candidates can be measured by the VLBI and EHT project~\cite{Fish,Lu,Psaltis}. The intensity map of the shadow image depends on the model of the accretion disk and the emission process, although the boundary of the shadow is completely determined by the geometry of spacetime. The shadow circle is slightly deformed for a rotating black hole. Both the circle for the Schwarzschild-MOG black hole, $a=0$, and the deformed circle for the Kerr-MOG black hole, $a\neq 0$, are significantly increased in size as $\alpha$ increases from zero. This can help determine whether Einstein's general relativity theory is correct for strong gravitational fields.

\section*{Acknowledgments}

I thank Martin Green and Viktor Toth for helpful discussions. I thank Viktor Toth for his help in calculating the black hole shadow images. Research at the Perimeter Institute for Theoretical Physics is supported by the Government of Canada through industry Canada and by the Province of Ontario through the Ministry of Research and Innovation (MRI).

\begin{figure} [h]
\centering \includegraphics[scale=0.6]{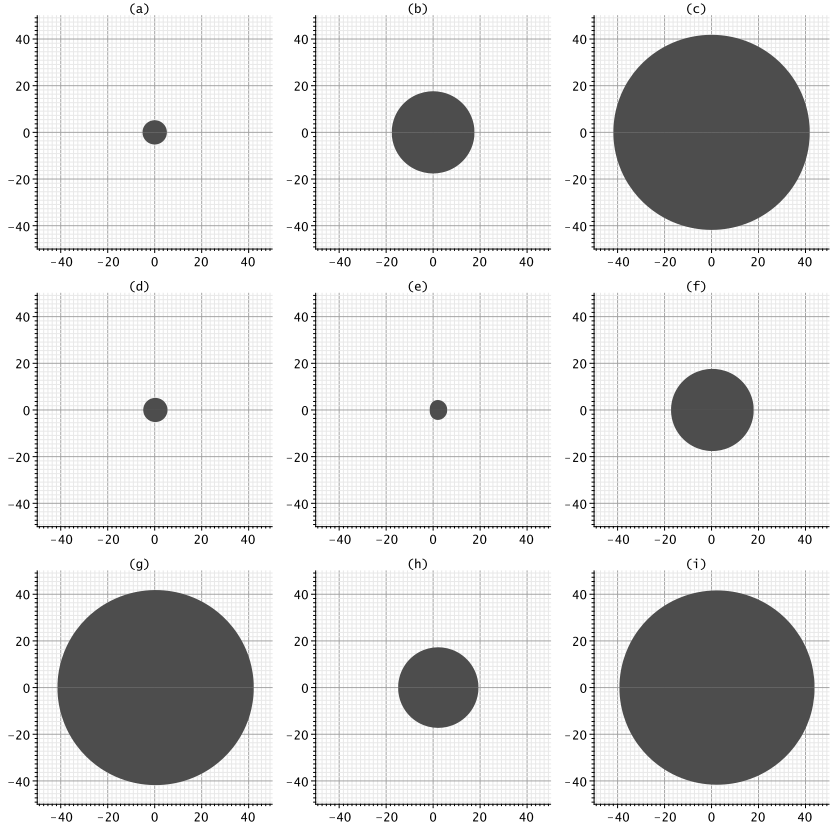}\\
\caption{\label{fig:BHshadow}(a) Black hole shadow for $G_N=1$, $M=1$, $a=0$, $\alpha=0$.
(b) Black hole shadow for $G_N=1$, $M=1$, $a=0$, $\alpha=3$. (c) Black hole shadow for $G_N=1$, $M=1$, $a=0$, $\alpha=9$.
(d) Black hole shadow for $G_N=1$, $M=1$, $a=0.16$, $\alpha=0$ and $\theta=63\,^{\circ}$. (e) Black hole shadow for $G_N=1$, $M=1$, $a=0.95$, $\alpha=0$ and $\theta=63\,^{\circ}$. (f) Black hole shadow for $G_N=1$, $M=1$, $a=0.16$, $\alpha=3$ and $\theta=63\,^{\circ}$. (g) Black hole shadow for $G_N=1$, $M=1$, $a=0.16$, $\alpha=9$ and $\theta=63\,^{\circ}$.
(h) Black hole shadow for $G_N=1$, $M=1$, $a=0.95$, $\alpha=3$ and $\theta=63\,^{\circ}$. (i) Black hole shadow for $G_N=1$, $M=1$, $a=0.95$, $\alpha=9$ and $\theta=63\,^{\circ}$.
}
\end{figure}

\end{document}